\title{\vspace{-2.0cm}\textbf{Quantifying the strength of structural connectivity underlying functional brain networks}}
\author{Phebe Brenne Kemmer$^{1}$, F. DuBois Bowman$^{2}$ and Helen Mayberg$^{3}$ and Ying Guo$^{1}$ \\
	    \\
	$^{1}$Department of Biostatistics and Bioinformatics,\\
	 Rollins School of Public Health, \\
	 Emory University, Atlanta, GA \\
	 \\
	$^{2}$Department of Biostatistics, \\
	Mailman School of Public Health, \\
	Columbia University, New York, NY\\
	\\
	$^{3}$Departments of Psychiatry and Neurology,\\
	Emory University School of Medicine, \\
	Atlanta, GA \\
	\\
	*\textit{email}: yguo2@emory.edu}

\documentclass[11pt]{article}
\usepackage{fullpage,amsmath,amssymb,graphicx,booktabs,multirow,indentfirst,sectsty,enumerate,outlines,enumitem,rotating,csquotes}
\usepackage{lscape}
\usepackage{graphicx}
\usepackage{tikz}
\usepackage{mathtools}
\usepackage{amssymb}
\usepackage{setspace}
\usepackage[round,sort]{natbib}
\usepackage[singlelinecheck=false]{caption}
\begin{document}
\maketitle

\singlespacing
\begin{abstract}
In recent years, there has been strong interest in neuroscience studies to investigate brain organization through networks of brain regions that demonstrate strong functional connectivity (FC). Several well-known functional networks have been consistently identified in both task-related and resting-state fMRI across different study populations. These networks are extracted from observed fMRI using data-driven analytic methods such as independent component analysis (ICA). A notable limitation of these FC methods is that they do not include or provide any information on the underlying structural connectivity (SC) which is believed to serve as the basis for interregional interactions in brain activity. We propose a new statistical measure of the strength of SC (sSC) underlying FC networks obtained from data-driven methods. The sSC is developed using information from diffusion tensor imaging (DTI) data. A key advantage of sSC is that it is a standardized coefficient which adjusts for the different number of voxels and baseline SC of various functional networks. Hence, sSC can be applied to compare the strength of structural connections across different FC networks. Furthermore, we propose a reliability index for data-driven FC networks to measure the reproducibility of the networks through re-sampling the observed data. By evaluating the association between the sSC and the reliability index, we can investigate whether underlying SC informs the reliability of identified FC networks. To perform statistical inference such as hypothesis testing on the sSC, we develop a formal variance estimator of sSC based a spatial semivariogram model with a novel distance metric. We demonstrate the performance of the sSC measure and its estimation and inference methods with simulation studies. For real data analysis, we apply our methods to a multimodal imaging study with resting-state fMRI and DTI data from 20 healthy controls and 20 subjects with major depressive disorder. Results show  that well-known resting state networks all demonstrate higher SC within the network as compared to the average structural connections across the brain. We also found that sSC is positively associated with the reliability index, indicating that the FC networks that have stronger underlying SC are more reproducible across samples. These results provide evidence that structural connections do serve as structural basis for the FC networks and that the structural information from DTI data can be leveraged to inform the reliability of functional networks derived through data-driven methods.\\
\end{abstract}

\newpage

%\doublespacing
%\setstretch{1.8}
%to add in references later, use \citep{Honey2009} for example
\section{Introduction}
%In-vivo neuroimaging techniques, such as functional magnetic resonance imaging (fMRI) and diffusion tensor imaging (DTI), have provided a pathway for researchers to study the function and structure of the human brain, respectively. Emerging evidence suggests that structural connections mediate functional connectivity (FC), by providing a mechanism of communication between spatially remote brain locations \citep{Collin2013,Hagmann2008,VanDenHeuvel2009,Toosy2004}, although FC can still exist between brain regions without direct structural links \citep{Honey2009}. Multimodal approaches that incorporate information from both fMRI and DTI can improve our understanding of the structure-function relationship in the brain, and allow us to characterize how  these connections are affected by brain diseases.\\
%\\
In recent years, network-oriented analysis has become a common approach in neuroimaging studies involving fMRI. A longstanding neurophysiologic principle holds that neural processing utilizes functionally specialized areas in the brain, which interact as components of highly complex networks. An important objective of many neuroimaging analyses is to spatially organize brain activity into distributed systems by evaluating the relatedness or correlations between spatially remote neurophysiologic events \citep{Friston1993}, also called functional connectivity (FC). The components of such systems consist of brain regions that exhibit a high degree of homogeneity within and larger variation between the networks. The identified functional networks may relate to resting-state brain activity or to neural activity associated with intrinsic neural processing, cognitive, emotional, visual, and motor functions \citep{Smith2009,Buckner2008}.\\
\\
In network-oriented analysis, the main objective is to evaluate functional connectivity (FC), defined as the temporal coherence between spatially remote neurophysiologic events \citep{Friston1993}. Numerous statistical approaches have been applied to evaluate FC. These methods can be generally grouped into two classes of approaches, namely correlation-based and partitioning methods. The correlation procedures quantify dependence between spatially distinct brain regions with correlations or partial correlations based on fMRI BOLD time series \citep{Hampson2002,Bowman2008,Wang2016}. The second class of data-driven FC approaches uses partitioning algorithms to identify spatially distinct components or clusters in the brain, with each component containing voxels exhibiting similar neural processing characteristics over time. For example, one of the first partitioning methods that has had applications in fMRI studies is cluster analysis \citep{Bowman2004,Bowman2004b,Cordes2002}. In recent years, independent component analysis (ICA) has become the most frequently used partitioning method to identify brain functional networks \citep{Beckmann2004,Beckmann2005a,Calhoun2001,Guo2008,Shi2016}.\\
\\
For the FC networks, it is commonly assumed that axonal fiber tracts serve as the basis for interregional interactions in brain activity. However, the nature of this structure-function relationship is only beginning to be revealed. In earlier work, evidence of the SC and FC relationship has emerged from computational work as well as clinical observations. For example, computational research has suggested that the underlying anatomical architecture of the cerebral cortex shapes resting-state FC on multiple timescales \citep{Honey2007,Ghosh2008}. Earlier clinical observations have revealed that interhemispheric FC is directly related to callosal integrity \citep{Quigley2003,Putnam2008}. Recent advances in diffusion MRI now provide a direct approach to empirically examine the structure-function relationship in human brains. In some recent work, joint analysis of fMRI and DTI data has provided evidence of a general correspondence between functional connectivity (measured by correlation-based FC) and structural connectivity (measured by diffusion tractography) \citep{Honey2009,Hagmann2008,Koch2002}. However, some important questions still remain to be addressed. First, whether and how the strength of structural connectivity varies across functional networks. This will help us to understand the difference in anatomical structure underlying various FC networks. Second, whether the strength of structural connectivity is associated with the reliability or robustness of the FC networks. Researchers have observed that some functional networks are more reproducible or robust across studies and samples. It will be interesting to investigate whether the difference in underlying SC plays a role in this. Third, whether the strength of structural connectivity underlying each of the FC networks differs between subpopulations, such as subjects with psychiatric or neurologic diseases vs. healthy controls.\\
\\
In addressing these questions, we need to develop a meaningful measure of SC that is comparable across networks and also across subjects. In existing work, raw measures of SC such as the count or proportion of fiber track connections derived from a probabilistic tractography algorithm based on DTI data are often used as the measure of strength of SC. These raw measures have several issues that limit their comparability across networks and across subjects: they don't account for the difference in the size or number of voxels across FC networks, nor do they account for the difference in the baseline structural connections at various brain locations. For example, some FC networks may consist of regions that are adjacent to high traffic white matter areas where there is a high density of fiber bundles. The raw SC measures will tend to be higher for these FC networks compared to others that are not located adjacent to the major fiber bundles, even if the other networks actually have stronger and more coherent within-network SC. Moreover, the raw SC measures have limited comparability across subjects due to the differences in baseline SC between subjects.\\
\\
In this paper, we present a novel measure of the strength of structural connectivity (sSC) underlying a functional network, which can help us understand the structural underpinnings of functional interactions in the brain. We aim to use this measure to characterize the sSC underlying FC networks estimated by data-driven FC method such as ICA. A key advantage of sSC is that it a standardized coefficient which adjusts for the different number of voxels and baseline SC related to a functional network. This makes sSC a more suitable measure to compare the strength of structural connections across different FC networks and to investigate differences in SC between subpopulations. We develop an estimator for sSC based on DTI probabilistic tractography. Furthermore, for statistical testing purposes, we develop a variance estimator to quantify the variability of the estimated sSC. The variance estimator is derived from a parametric semivariogram model with a novel distance metric. Based on the proposed inference framework, we develop three statistical tests for addressing various questions in FC and SC analysis. The first test evaluates the significance of the sSC of a given functional network. The second test examine the difference in SC between functional networks. The third test evaluates whether there is significant difference in sSC between subject groups. We conduct simulation studies to evaluate the performance of our proposed measure, and apply our method to a resting-state fMRI and DTI dataset.\\
\\
We can use the proposed sSC measure to investigate whether underlying structure connectivity informs the reliability of FC networks derived from data-driven methods. One issue of data-driven FC analysis is that the estimated FC networks often vary across samples or methods.  For example, the results from an ICA run may vary based on the choice of algorithm, starting values, subject variability, or data preprocessing steps \citep{Calhoun2004}. Some functional networks may not be reproducible in different analyses or in other data sets. Determining the reliability of the estimated ICs obtained from a given data set is important for appropriate interpretations of the extracted networks. Previous work has proposed assessing reliability of the estimated networks by evaluating the reproducibility in repeated ICA runs, with different initial conditions, with data resampled from the original functional time series, or with simulated data having a known brain network structure \citep{Himberg2004,Meinecke2002,Duann2006}. These approaches measure the reliability of networks only through the algorithmic and statistical reproducibility of ICA applied to fMRI data. The goal in the current paper is to evaluate whether the structural connections underlying these FC networks are related to the reliability of these networks. To this end, we propose a reliability index that measures the reproducibility of the estimated functional networks based on bootstrapped samples. We then evaluate the association between the sSC and the reliability index across functional networks.\\
%Finally, we present a new FC-SC reliability (FSR) index that considers both the statistical reproducibility and the strength of the underlying SC. This measure will allow us to distinguish different types of FC networks such as those which have both high FC and SC versus those with high FC but low SC.
\\
Several well-known functional networks have been consistently identified in both task and resting-state fMRI studies across different populations \citep{Smith2009,Buckner2008,Damoiseaux2006,Laird2011}. Some networks are identified more robustly than others, although it is unclear why. Our proposed measure of sSC will allow us to investigate the underlying SC of different functional networks, and better understand the differences in their anatomical underpinnings, reliability, and other aspects of their nature. %As the relationship between brain structure and function unfolds, it is important to develop statistical methods that combine information from fMRI and DTI data.
\section{Data}
%DuBois' suggestion: Introduce the data here (description is currently in Data application section), and make reference to the data in the methods to make the ideas more concrete to the reader. \\
%%%%%%%%%%%%%%%%%%%%%%%%%%%%%%%%%%%%%%%%%%%%%%%%%%%%%%%
%%%%%%%%%%%%%%%%%%%%%%%%%%%%%%%%%%%%%%%%%%%%%%%%%%%%%%%
\subsection{Subjects}
We estimated the FC networks from 20 healthy subjects. They had an average age of 42.4 years (SD: 9.0) and were 50\% male. Using the resting-state fMRI (rs-fMRI) and DTI data collected from this group of subjects, we used the proposed sSC measure to evaluate the SC underlying the extracted FC networks and investigated the association between sSC and the reliability measure. In addition, we compared the sSC between these 20 healthy subjects vs. 20 patients with major depressive disorder (MDD). The MDD patients were matched with the healthy control subjects by age and gender, with an average age of 45.8 years (SD: 9.6) and 50\% male. The mean Hamilton Depression Rating Scale (HAM-D) score for the MDD subjects was 19 (SD: 3.4), indicating "Severe Depression" \citep{Hamilton1960}. The average length of their current depression episode was 82 weeks.
%At the time that the scans were collected, all MDD patients had unipolar depression and had never been treated. During the scan, subjects were instructed to lie passively in the scanner and to refrain from thinking about anything in particular. Since resting-state captures the "idling" brain, and the brains of MDD patients seem to be "stuck" in an idle state, this paradigm is well-suited for studying this disorder. \\
\\
%\indent MDD affects approximately 1 in 6 people in their lifetimes, causes substantial occupational and social impairments, and is co-morbid with anxiety and substance abuse disorders \citep{Kessler2003}. Although there have been many resting-state studies conducted on MDD, they do not all agree about the mechanisms of connectivity disruption, and its pathology is still unclear \citep{Northoff2011}. Functional and structural neuroimaging play critical roles in advancing our knowledge about this disorder. \\
%%%%%%%%%%%%%%%%%%%%%%%%%%%%%%%%%%%%%%%%%%%%%%%%%%%%%%
%%%%%%%%%%%%%%%%%%%%%%%%%%%%%%%%%%%%%%%%%%%%%%%%%%%%%%
\subsection{Data acquisition and preprocessing}
\indent DTI, rs-fMRI, and T1-weighted data were collected in a single session with 3T Siemens Tim Trio scanner. The DTI sequence consisted of 60 scans with different diffusion-weighted directions ($b=1000s/mm^2$) and four non-diffusion weighted scans ($b=0$), acquired using a single-shot spin-echo echo planar imaging (EPI) sequence. Additional DTI scanning parameters include: TR=11300 ms, TE=104 ms, GRAPPA on, FOV=256 mm, number of slices=64, voxel size=2x2x2 mm, and matrix size=128x128. For registration purposes, high-resolution T1-weighted images were collected using a 3D MPRAGE sequence with the following parameters: TR=2600 ms, TI=1100 ms, TE=3 ms, number of slices = 176, voxel size = 1×1×1 mm, matrix size= 224×256, flip angle=$8^{\circ}$. Functional images were collected over 150 time points, with a z-saga sequence to minimize artifacts in the medial PFC and OFC due to sinus cavities. rs-fMRI scans were acquired interleaved with the following parameters: TR=2.92 seconds; TE1=30 ms, TE2=66 ms, flip angle=$90^{\circ}$, number of axial slices=30, slice thickness=4 mm, FOV=220 mm, and total duration=7.3 min. \\
\\
Several standard preprocessing steps were applied to the rs-fMRI data, including despiking, slice timing correction, motion correction, registration to MNI 2mm standard space, normalization to percent signal change, removal of linear trend, regressing out CSF, WM, and 6 movement parameters, bandpass filtering (0.009 to 0.08), and spatial smoothing with a 6mm FWHM Gaussian kernel. Preprocessing steps for the DTI data include brain extraction to remove non-brain regions, phase reversal distortion correction, and aligning diffusion weighted images to the average non-diffusion weighted image by rigid body affine transformation to remove motion and eddy-current induced artifact. We then estimate the directional diffusion at each voxel based on a diffusion tensor model implemented via the Diffusion Toolbox (FDT) in FSL \citep{Behrens2003}.
%%%%%%%%%%%%%%%%%%%%%%%%%%%%%%%%%%%%%%%%%%%%%%%%%%%%%%
%%%%%%%%%%%%%%%%%%%%%%%%%%%%%%%%%%%%%%%%%%%%%%%%%%%%%%
\subsection{Functional Network Estimation}
We estimated group-level functional networks from the rs-fMRI data using a spatial ICA implemented in Matlab using the Group ICA for fMRI Toolbox (GIFT) \citep{Calhoun2001}. We identified the set of components that represent functional networks by comparing the extracted component maps with those of well-established resting state functional networks \citep{Smith2009}.

%After extracted FC networks based on the ICA were The group ICA yields a common set of spatially independent component (IC) maps and their associated time series. Many of the IC maps correspond to known resting state networks, such as the default mode network \citep{Buckner2008}, while others represent noise and artifacts. To determine which components reflect true FC networks, we evaluate their correlation with the set of major resting-state network maps defined by \citet{Smith2009}; we use the cutoff of $r \ge 0.25$ to determine whether a component corresponds to a known network.
%%%%%%%%%%%%%%%%%%%%%%%%%%%%%%%%%%%%%%%%%%%%%%%%%%%%%%%
%%%%%%%%%%%%%%%%%%%%%%%%%%%%%%%%%%%%%%%%%%%%%%%%%%%%%%%
\subsection{Structural Connectivity Estimation}
Structural connectivity is assessed using DTI data, a diffusion-weighted MRI technique that estimates the location of structural fibers in the brain by measuring the direction and magnitude of water diffusion, since water tends to diffuse along white matter fiber tracts \citep{Johansen-Berg2009}. We evaluated structural connectivity across the whole brain by implementing a widely-used probabilistic tractography approach via FSL's diffusion toolbox \citep{Behrens2007,Behrens2003}.\\
\\
This procedure successively initiates streams, which are intended to trace the paths of white matter fibers in the brain. A specified number of streams (5000 in our analysis) are sent from a seed voxel and traced voxel-by-voxel as they pass through the diffusion tensor field, terminating according to a stopping rule. This procedure yields a voxel-level count (out of 5000 trials) which empiricially quantifies the probability of SC between the seed voxel and each other target voxel in the brain \citep{Behrens2007,Johansen-Berg2009}. To evaluate the sSC underlying each functional network, we use the white matter voxels in each corresponding independent component as the seed locations for this procedure.
%%%%%%%%%%%%%%%%%%%%%%%%%%%%%%%%%%%%%%%%%%%%%%%%%%%%%%%%%%%
%%%%%%%%%%%%%%%%%%%%%%%%%%%%%%%%%%%%%%%%%%%%%%%%%%%%%%%%%%%
\section{Methods}\label{methods}
\subsection{The strength of Structural Connectivity (sSC) measure}
In this section, we propose a new statistical measure for quantifying the strength of structural connectivity (sSC) underlying functional networks. This measure can help compare the SC across different functional networks. Suppose that based on a data-driven FC method such as group ICA, we extracted $q$ functional networks of interests.  In group ICA, the spatial distribution of the functional networks are reflected by the component maps estimated from ICA.  We propose the following measure, $\theta_\ell$ to quantify the sSC underlying the functional network represented by component $\ell$  ($\ell=1,...,q$),
\begin{equation}\label{sSC-def}
	\theta_{\ell} =\frac{\displaystyle\sum_{\substack{j,k \in \Omega_{\ell}}}[p_{jk} -(\bar{p_j}+\bar{p_k})/2]}{\displaystyle\sum_{j,k \in \Omega_{\ell}}[1 -(\bar{p_j}+\bar{p_k})/2]}.
\end{equation}
\\
%Interpretation of $\theta_{\ell}$\\
Here, $\Omega_{\ell}$ is the set of voxels within the $\ell^{th}$ IC; $p_{jk}$ is the probability of SC between the pair of voxels $j$ and $k$ within component $\ell$; $\bar{p_j}$ is the average probability of structural connection between voxel $j$ and the rest of the brain, which is defined as follows,\\
\begin{center}$\bar{p}_j=\dfrac{1}{V-1}\displaystyle\sum_{\substack{v=1 \\ v \ne j}}^{V}{p_{jv}}$ , \\
\end{center}
where V is the total number of voxels within the brain. Similarly,  $\bar{p_k}$ is defined analogously to represent the average probability of structural connection between voxel $k$ and the rest of the brain.\\
\\
In the definition of sSC, $p_{jk}$ is the raw probability of structural connectivity between the voxel pair.  The raw SC probability may be affected by factors such as the location of the voxels in the brain and the average strength of structural connections across the brain, which limits the comparability of the raw SC probability across functional networks and subjects. To help address this issue, we adjust the raw SC probability by the baseline SC strength, i.e. $(\bar{p_j}+\bar{p_k})/2$, which represents the average overall probability of SC between the voxels $j$, $k$ in network $\ell$ and a random location in the brain.  This adjustment accounts for the baseline SC that is related to the locations of the voxels and the average SC across the brain. Thus, the numerator of $\theta_{\ell}$ reflects the degree to which the raw SC within the functional network exceeds the baseline SC between the network and the rest of the brain expected on average. This adjustment allows us to  compare the underlying SC of networks located in high-traffic vs. low-traffic SC areas and also improve the comparability of sSC across subjects who may be associated with different baseline SC.
%voxel $j$ located in a hub structurally connected to several brain areas would yield a high $\bar{p}_j$, whereas another voxel $j'$ in a low-traffic area of the brain would yield a low $\bar{p}_{j'}$.\\
We then further standardize the sSC measure by dividing by the maximum possible value, in which there is complete SC between all voxel pairs within the functional network (i.e. $p_{jk}=1$). The sSC is then within the standard scale of less than or equal to 1, which further facilities the comparison of sSC across functional networks with various sizes (i.e. total number of voxels).  The definition of the sSC measure is related to that of the Kappa coefficient \citep{Cohen1968}, in the sense that sSC represents the observed strength of SC, relative to the baseline SC expected by chance, divided by the maximum possible value.\\
\\
In the sSC measure definition in (\ref{sSC-def}), $j$ and $k$ represent voxels in a functional network. Hence, the proposed sSC is a voxel-level measure of SC underlying an FC. Alternatively, we can define sSC in the similar way on the region-level, where $j$ and $k$  represent regions within an functional network.

\subsection{Estimation and Inference for the sSC measure}
\subsubsection{Estimate of the sSC measure}
In this section, we present statistical estimation and inference methodology for the proposed sSC measure. We estimate the probability of SC between brain locations using a probabilistic tractography procedure with DTI data \citep{Behrens2007}, as described in section 2.4. This procedure initiates $N$ streams from seed voxel $j$ and tracks how many of these streams pass through target voxel $k$; the number of streams connecting these voxels is denoted by $N_{jk}$. The probability of SC for voxel pair $j,k$, $p_{jk}$, is thus estimated as  $\hat{p}_{jk}=N_{jk}/N$.
\\
Plugging in the estimated probability of SC derived from probabilistic tractography yields the following estimator for the sSC measure of the $\ell$th functional network:\\
\begin{equation}
	\hat{\theta}_{\ell} =\frac{\displaystyle\sum_{\substack{j,k\in\Omega_{\ell}}}[N_{jk} -(\bar{N_j}+\bar{N_k})/2]}{\displaystyle\sum_{j,k\in\Omega_{\ell}}[N -(\bar{N_j}+\bar{N_k})/2]},
\end{equation}
where $N_{jk}$ is the number of streams passing through voxels $j$ and $k$, $\bar{N_j}=\frac{1}{V-1}\sum_{v=1,v\ne j}^{V} N_{jv}$ is the average streams connecting voxel $j$ with the rest of the brain, $\bar{N_k}=\frac{1}{V-1}\sum_{v=1,v\ne k}^{V} N_{kv}$ is the average streams connecting voxel $k$ with the rest of the brain, and $N$ is the total number of streams initiated from each voxel in the probabilistic tractography procedure. \\
\\
%region-level vs voxel-level
We have mentioned that the sSC measure can also be defined on the region-level, where $j$ and $k$ represent a pair of regions rather than voxels within an FC network. The estimation of the region-level sSC is similar to that of the voxel-level sSC. The main difference is that $N_jk$ would then represent the number of streams connecting two region $j$ and $k$ in a probabilistic tractography procedure. We note that when the sSC measure is defined on the voxel-level, the magnitude of $\hat{\theta}_{\ell}$ will inherently be small relative to its upper bound of 1, because the probability of SC between two individual voxels tends to be very low due to the small target size. The magnitude of the estimate for the region-level sSC is expected to be larger because the probability of SC between two regions are higher due to multiple voxels contained in each region.\\
\\
\subsubsection{Variance of the sSC estimate}
\indent In order to conduct statistical inference such as hypothesis testing for $\theta_{\ell}$, we propose a variance estimator for quantifying the variability in the estimated sSC, i.e. $var(\hat{\theta}_{\ell})$. We note that the sSC estimator $\hat{\theta}_{\ell}$ is a function of the observed stream counts between pairwise voxels from a probabilistic tractography procedure. To facilitate deriving the variance of $\hat{\theta}_{\ell}$, we define $\boldsymbol{ N^*}$ as a vector which contains the stream counts $N_{jk}$ for all voxel pairs $\{j,k\}$, i.e.
%where at least one of the voxel is within the $\ell$th FC network, i.e. $\mathbb{P}_\ell=\{N_j,j', \textrm{s.t.} j \textrm{or} j'\in\Omega_{\ell}\}$. It can be shown that $\|\mathbb{P}_\ell \|=$ \begin{center}
	$\boldsymbol{ N^*}=\begin{bmatrix}
	\{N_{jk}\}
	\end{bmatrix}=\begin{bmatrix}
	N_{12}\\
	N_{13}\\
	\vdots \\
	N_{V-1,V}
	\end{bmatrix}$ \\
%\end{center}

\noindent  We can show that the proposed sSC estimate, $\hat{\theta}_{\ell}$, can be written as a function of $\boldsymbol{ N^*}$ as follows (see Appendix B):
\begin{equation}
\label{theta_rewrite}
	\hat{\theta}_{\ell}=\dfrac{(\boldsymbol{C_{\ell}}-\boldsymbol{A})\boldsymbol{N^*}}{ b-\boldsymbol{AN^*}},
\end{equation}	
%\begin{center}
%	$\hat{\theta}_{\ell}=\dfrac{\left(\boldsymbol{C_{\ell}}-\dfrac{(V_{\ell}-1)}{2(V-1)} \displaystyle\sum_{\substack{j\in\Omega_{\ell}}}\boldsymbol{C_j}\right)\boldsymbol{N^*}}{ \dfrac{V_{\ell}(V_{\ell}-1)}{2}N-\left(\dfrac{(V_{\ell}-1)}{2(V-1)} \displaystyle\sum_{\substack{j\in\Omega_{\ell}}}\boldsymbol{C_j}\right)\boldsymbol{N^*}}$
%\end{center}
\vskip10pt
\noindent where \hskip5pt $\boldsymbol{A}=\dfrac{(V_{\ell}-1)}{2(V-1)} \displaystyle\sum_{\substack{j\in\Omega_{\ell}}}\boldsymbol{C_j}$ \hskip10pt and \hskip10pt $b=\dfrac{V_{\ell}(V_{\ell}-1)}{2}N$. \\

\noindent Here, $\boldsymbol{C_{\ell}}$ and $\boldsymbol{C_j}$ are $1 \times {V \choose 2}$ row vectors of binary indicators. The $z$th $(z=1,\ldots,{V \choose 2})$ element of $\boldsymbol{C_{\ell}}$ equals 1 if the $z$th voxel pair in $\boldsymbol{ N^*}$ both belong to $\Omega_{\ell}$, and 0 otherwise; the $z$th $(z=1,\ldots,{V \choose 2})$ element of $\boldsymbol{C_j}$ is 1 if the $z$th voxel pair in $\boldsymbol{ N^*}$ involves voxel $j$, and 0 otherwise, $V$ is the total number of voxels in the whole brain, and $V_{\ell}$ is the number of voxels in the $\ell$th network. \\
%$N$ is the maximum possible number of connections between a pair of voxels (i.e. the number of streams initiated from each voxel in a probabilistic tractography procedure).
\\
Equation (\ref{theta_rewrite}) shows that the numerator and denominator of $\hat{\theta}_{\ell}$ are each a linear function of the random vector $var(\boldsymbol{ N^*})$ and some constants. Therefore, the variance of $\hat{\theta}_{\ell}$ can be obtained from $var(\boldsymbol{ N^*})$ using the Delta method \citep{Casella1990}. In the following, we provide an approach for deriving $var(\boldsymbol{ N^*})$. First, we note that each element $N_{jk}$ in $\boldsymbol{ N^*}$ is an binomial count of the number of streams passing the voxel pair $\{j,k\}$ out of a total of $N$ initiations. Therefore,  $N_{jk}$ follows a binomial distribution with $N_{jk} \sim \text{Bin}(N,p_{jk})$. Given that $N$ is large, the binomial distribution can approximated with the Normal distribution, $N_{jk} \sim \text{N}(\mu_{jk},\sigma^2_{jk})$ where $\mu_{jk}=Np_{jk}$ and $\sigma^2_{jk}=Np_{jk}(1-p_{jk})$.   The vector $\boldsymbol{ N^*}$ approximately follows a multivariate normal distribution, i.e. $\boldsymbol{ N^*} \sim MVN(\boldsymbol{\mu},\boldsymbol{\Sigma})$ where
%The mean vector $\boldsymbol{\mu}$ is a ${V \choose 2}\times 1$ vector which includes $\{\mu_{jk}\}$ for all voxel pairs $\{j,k\}$ and $\boldsymbol{\Sigma}$ is the  ${V \choose 2}\times{V \choose 2}$ variance-covariance matrix. \\

%\indent That is, $\boldsymbol{N^*} \sim \text{MVN}(\boldsymbol{\mu},\boldsymbol{\Sigma})$, where\\
\begin{center}
	$\boldsymbol{\mu}=\begin{bmatrix}
	\{\mu_{jk}\}
	\end{bmatrix}=N\begin{bmatrix}
	p_{12}\\
	p_{13}\\
	\vdots \\
	p_{V-1,V}
	\end{bmatrix}, \textrm{  and}$
\end{center}
\vskip5pt
\begin{center}
	$\boldsymbol{\Sigma}=\begin{bmatrix}
	\{cov(N_{jk},N_{j'k'})\}
	\end{bmatrix}
	=\begin{bmatrix}
	var(N_{12}) & cov(N_{12},N_{13}) & \cdots & cov(N_{12},N_{V-1,V})\\
	cov(N_{13},N_{12}) & var(N_{13}) & \cdots & cov(N_{13},N_{V-1,V})\\	
	\vdots & \vdots & \ddots & \vdots \\
	cov(N_{V-1,V},N_{12}) & cov(N_{V-1,V},N_{13}) & \cdots & var(N_{V-1,V})
	\end{bmatrix}$ \\
\end{center}

We have shown the diagonal variance elements of the variance-covariance matrix $\boldsymbol{\Sigma}$ are $var(N_{jk})=\sigma^2_{jk}=Np_{jk}(1-p_{jk})$. The derivation of the the off-diagonal covariance elements are more challenging. Given the high dimensionality of $\boldsymbol{\Sigma}$, there are a large number of covariance terms. Furthermore,  deriving the covariance between pairs of streams counts $N_{jk}$ and $N_{j'k'}$ is not straightforward because of the potential spatial dependence between them.  In this paper, we propose a novel approach based on a parametric semivariogram model \citep{Cressie1993} to estimate the covariance between pairs of stream counts from a probabilistic tractography procedure.  Specifically, we model each covariance term, $cov(N_{jk},N_{j'k'})$, as a function of the distance between voxel pair $j,k$ and voxel pair $j',k'$, denoted $d_{jk,j'k'}$, which decays as the distance between the observations increases. One distinct feature in modeling spatial dependence between the tractography stream counts is that there is no straightforward way to define the spatial distance between $N_{jk}$ and $N_{j'k'}$.  Unlike the standard case, these stream counts are not a measurement at a single spatial location but rather an outcome observed between a pair of spatial locations, i.e. voxels.  In other words, we need a metric that quantifies the spatial distance between \textit{pairs} of voxel pairs. For this purpose, we propose a novel distance metric between two voxel pairs $(j,k)$ and $(j',k')$, \\
\begin{equation}
\label{dist_func}
	d_{jk,j'k'}=\min\left[\dfrac{d^{\ast}(j,j')+d^{\ast}(k,k')}{2},\dfrac{d^{\ast}(j,k')+d^{\ast}(k,j')}{2}\right],
\end{equation}
where $d^{\ast}(j,k)$ is the euclidean distance between voxels $j$ and $k$ which serves as a proxy for the distance between voxels. This distance metric is based on the average euclidean distances for each pairwise combination of voxels between the first voxel pair $j,k$ and the second voxel pair $j',k'$. This metric captures the spatial dependence between the pair of voxel pairs.\\
\\
%The exponential semivariogram model is defined as:
%$\gamma(h;\theta)=c_0+(c_0-c_n)[1-\exp(-h/a_0)], h>0$
\\
The semivariogram offers a flexible approach to model the covariance as a function of spatial distance, since there are many different types of semivariogram models (e.g. exponential, Gaussian, spherical, Matern class, etc.), each of which is characterized by shape parameters known as the range, sill, and the nugget effect. Thus, we can use the semivariogram model to incorporate spatial dependence (via our proposed distance metric (\ref{dist_func}) in the estimation of the covariance between stream counts. The model type and its associated parameters can then be estimated by fitting to the empirical semivariogram using the observed data \citep{Cressie1993}.\\
\\
After obtaining the variance-covariance matrix $\boldsymbol{\Sigma}$ for $\boldsymbol{ N^*}$ ,we can then derive the variance of $\hat{\theta}_{\ell}$ as follows using the Delta method.  % (see Appendix B for details).
\begin{eqnarray}\label{var_est}
  var(\hat{\theta}_{\ell}) &=& var \left( \dfrac{(\boldsymbol{C_{\ell}}-\boldsymbol{A})\boldsymbol{N^*}}{b-\boldsymbol{AN^*}} \right) \nonumber \\
& \approx & \left[\dfrac{(\boldsymbol{C_{\ell}}-\boldsymbol{A})\boldsymbol{\mu}}{b-\boldsymbol{A\mu}} \right]^2
\left[ \dfrac{(\boldsymbol{C_{\ell}}-\boldsymbol{A})\boldsymbol{\Sigma}(\boldsymbol{C_{\ell}}-\boldsymbol{A})'}{[(\boldsymbol{C_{\ell}}-\boldsymbol{A})\boldsymbol{\mu}]^2}
+ \dfrac{\boldsymbol{A\Sigma A'}}{[b-\boldsymbol{A\mu}]^2}
- 2 \dfrac{[-(\boldsymbol{C_{\ell}}-\boldsymbol{A})\boldsymbol{\Sigma A'}}{[(\boldsymbol{C_{\ell}}-\boldsymbol{A})\boldsymbol{\mu}][b-\boldsymbol{A\mu}]}\right]
\end{eqnarray}
\\
%The semivariogram is a standard statistical measure of spatial variability that incorporates the distance between observations \citep{Minasny2005}.

In addition to the parametric variance estimator proposed in (\ref{var_est}), we will also consider a nonparametric variance estimate based on the bootstrap method. We will compare the performance of the parametric vs. bootstrap variance estimators in simulation studies.
%The covariance function $C(h)$ is related to the semivariogram function $\gamma(h)$ by this equation: $\gamma(h)=C(0)-C(h)$.
% Then, because the numerator and denominator of $\hat{\theta}_\ell$ are linear combinations of $\boldsymbol{N^*}$, we can use the Delta method based on $\boldsymbol{\Sigma}$ to derive the approximate the large sample variance of $\hat{\theta}_{\ell}$, to be used in inference. %Although $var(\hat{\theta}_{\ell})$ is a scalar value, it requires estimation of the large ${V \choose 2}\times{V \choose 2}$ matrix $\boldsymbol{\hat{\Sigma}}$.
%%%
\subsubsection{Hypothesis testing based on the sSC measure}
Based on the estimated sSC $\hat{\theta}_\ell$ and its variance estimator, we can conduct statistical tests to compare the strength of SC across various functional networks and between groups of subjects. In this section, we present three hypothesis tests for 1) testing the significance of sSC within a functional network, 2) comparing sSC between two functional networks, and 3) comparing the sSC underlying a given functional network between subject subgroups.\\
\\
\indent In the first hypothesis test, we aim to test the significance of the strength of SC within a given functional network. For a network represented by the $\ell$th component from the ICA, $\theta_\ell=0$ indicates that observed strength of SC within the network is no higher than the average SC between the network and the rest of the brain. If an estimated component is representative of a true functional network, we expect the strength of SC within the network to be significantly above the average SC, i.e. $\theta_\ell > 0$. Thus, we can test the significance of the sSC within component $\ell$ by specifying the hypotheses as: $H_0: \theta_{\ell}=0$ vs. $H_1: \theta_{\ell}>0$. Suppose $\hat{\theta}_{i\ell}$ represent the estimated sSC from the $i$th subject in the data ($i=1,\ldots,n$), we propose the following Wald-type test statistic: \\
\begin{equation}\label{hyp1}
	Z^*=\dfrac{\bar{\hat{\theta}}_{\ell}}{\sqrt{\hat{Var}(\hat{\theta}_{\ell})/n}},
\end{equation}
where $\bar{\hat{\theta}}_{\ell}=\sum_{i=1}^{n}\hat{\theta}_{i\ell}/n$ and $\hat{Var}(\hat{\theta}_{\ell})$.  If $Z^*>z_{1-\alpha}$ where $z_{1-\alpha}$ is the $1-\alpha$ percentile of the standard normal distribution, we would reject the null hypothesis $H_0$ in favor of the alternative hypothesis $H_1$ and conclude there is significant sSC within the network characterized by the $\ell$th IC. Otherwise, we fail to reject $H_0$.\\
\\
The second type of hypothesis test aims to compare the strength of SC between two networks characterized by components $\ell$ and $\ell'$. The hypotheses in this case are $H_0: \theta_{\ell}=\theta_{\ell'}$ vs. $H_1: \theta_{\ell} \ne \theta_{\ell'}$. The test statistic is $\bar{\hat{\theta}}_{\ell}-\bar{\hat{\theta}}_{\ell'}$, where $\bar{\hat{\theta}}_{\ell}$ and $\bar{\hat{\theta}}_{\ell'}$ are the average sSC across subjects for the two networks. We can derive the p-value for this test statistic using a non-parametric permutation testing approach. Specifically, we permute the network label within each subject to generate an empirical distribution for the test statistic $\bar{\hat{\theta}}_{\ell}-\bar{\hat{\theta}}_{\ell'}$ .\\
\\
Thirdly, we can test whether the strength of SC for a given network, e.g. the $\ell$th network, differs between subject subgroups with the hypotheses: $H_0: \theta_{\ell,1}=\theta_{\ell,2}$ vs. $H_1: \theta_{\ell,1} \ne \theta_{\ell,2}$ where $\theta_{\ell,j}$ is the sSC underlying the $\ell$th network for the $j$th subgroup, $j=1,2$. We can test the group difference using the following test statistic,
\begin{equation}
	Z^*=\dfrac{\bar{\hat{\theta}}_{\ell,1}-\bar{\hat{\theta}}_{\ell,2}}{\sqrt{\dfrac{\hat{Var}(\hat{\theta}_{\ell,1})}{n_1}+\dfrac{\hat{Var}(\hat{\theta}_{\ell,2})}{n_2}}}
\end{equation}
Here, $n_1$ and $n_2$ are the number of subjects in each subgroup, and the definitions for the subgroup mean and variance terms are analogous to those in equation (\ref{hyp1}). If $|Z^*|>z_{1-\alpha/2}$ where $z_{1-\alpha/2}$ is the $1-\alpha/2$ percentile of the standard normal distribution, we would reject the null hypothesis $H_0$ in favor of the alternative hypothesis $H_1$ and conclude there is significant difference in the sSC between the two groups. Otherwise, we fail to reject $H_0$.
Alternatively, we can test the between-group hypotheses using a non-parametric permutation testing approach, in which we permute subject group label to generate an empirical distribution for the between-group difference $\bar{\hat{\theta}}_{\ell,1}-\bar{\hat{\theta}}_{\ell,2}$.
\subsection{Does sSC inform the reliability of functional networks?}
It is often of interest to investigate the reliability of functional networks derived from data-driven method such as ICA. The reproducibility of an estimated network varies across different runs of the algorithm or different samples of subjects. For example, due to the stochastic nature of the ICA algorithm and individual subject variability, the ICA-derived functional networks often vary across different fMRI studies or even across various ICA runs based on the same data. Researchers also noticed that some networks tend to be more reproducible compared to others. In this paper, we would like to investigate whether the strength of structural connections underlying functional networks is associated with their reliability. One hypothesis is that networks that have stronger direct structural connections may tend to be more reproducible given their underlying structural. In this section, we plan to investigate whether the SC is associated with the reproducibility of data-driven FC networks based on ICA. First, we need a measure to quantify the reproducibility of various functional networks. Generally, a more reproducible functional network is interpreted as a network that is more consistently identified with different subject data sets sampled from the population and with different initial computational conditions. Based on this, we propose a novel reliability index based on the following procedure. Basically, the index assesses how reproducible each of the original components is in bootstrap samples from the original data.\\
\\
Step 1:	generate $B$ bootstrap samples each containing $n$  subjects sampled with replacement from the original data set, and perform the group ICA to extract $q$ components in each bootstrap sample.\\
\\
Step 2: For the $\ell$th component extracted from the original data, we identify the corresponding component $\ell_b$ in the $b^{th}$ bootstrap sample by identifying the component from the bootstrap sample that shows the highest spatial correlation with the $\ell$th component from the original data.\\
\\
Step 3: For the $\ell$th ($\ell=1,\ldots,q$) component in the original data, we propose the following reliability index

\begin{equation}\label{rel-def}
	\large{R_{\ell}=\dfrac{\frac{1}{B} \sum_{b=1}^B |{r_{\ell \ell_b}}| - \frac{1}{Bq} \sum_{b=1}^B \sum_{\ell_{b}^*=1}^q|{r_{\ell \ell_b^*}}|}{1 - \frac{1}{Bq} \sum_{b=1}^B \sum_{\ell_{b}^*=1}^q|{r_{\ell \ell_b^*}}|}}
\end{equation}
where $r_{\ell\ell_b}$ is the spatial correlation between original component $\ell$ and its corresponding bootstrap component $\ell_b$, and $r_{\ell\ell_b^*}$ is the spatial correlation between component $\ell$ and  the $\ell_b^*$ component in the $b^{th}$ bootstrap sample.\\
% To help understand the reliability of various networks, it is important to determine the reliability of the estimated ICs to make accurate interpretations. To investigate whether strength of SC can be leveraged to inform IC reproducibility, we propose a novel reliability index based on bootstrapped ICA runs. \\
%We hypothesize that we can leverage SC information from DTI to inform the reliability of the estimated functional networks. To test our hypothesis, we will evaluate the association between the strength of SC (measured by $\hat{\theta}_{\ell}$) and IC reproducibility, as measured by a novel reliability index based on bootstrapped ICA runs. \\
\\
In the proposed reliability index, the first term in the numerator of $R_\ell$ represents the observed similarity between component $\ell$ and its corresponding component $\ell_b$ in the bootstrap samples. This term reflects how reproducible the $\ell$th component is across these bootstrap samples. Since component $\ell$ can be correlated with any bootstrap components by chance, even they represent different networks, we evaluate by-chance correlation between the $\ell$ th component and all the components from the bootstrap samples. Then we remove the by-chance correlation from the observed similarity and hence the numerator of $R_\ell$ represents the beyond-chance similarity between the $\ell_b$ th original component and its counterparts in bootstrap samples. We further standardize the measure by dividing by its maximum possible value, in which the original component $\ell$ is perfectly reproduced in the bootstrap sample (i.e. $r_{\ell\ell_b}=1$). Therefore, the proposed reliability index $R_\ell$ represents the observed reproducibility of a component $\ell$, corrected for the reproducibility expected by chance, and standardized by its maximum possible value. The reliability index $R_\ell$ ranges from 0 to 1, where $R_{\ell}=0$ indicates the $\ell$th IC is not reproducible in bootstrap samples after we correct for by-chance correlations across ICs and $R_{\ell}$ close to 1 indicates that the component is highly reproducible.

%Similar to the construction of the sSC measure $\theta_\ell$, $R_\ell$ is structured like the Kappa statistic, in that it represents the observed reproducibility of IC $\ell$, corrected for the reproducibility expected by chance, and divided by the maximum possible value.
%%%%%%%%%%%%%%%%%%%%%%%%%%%%%%%%%%%%%%%%%%%%%%%%%%%%%%%%
%%%%%%%%%%%%%%%%%%%%%%%%%%%%%%%%%%%%%%%%%%%%%%%%%%%%%%%%
\section{Simulation Studies}
%We conducted simulation studies to evaluate the performance of the proposed sSC statistic estimation and inference methods. We generated fMRI data for $n=20, 50$ subjects with $q=2$ underlying spatial sources. The spatial source maps are common across subjects and consist of one axial image slice of $10 \times 10$ voxels, for a total of $V=100$ voxels.
%(...How group-level maps are created...). After the IC maps are created, their temporal responses are generated based on real fMRI time series data of length $T=200$. Then Gaussian background noise with a variance of 1 (?? is this what I did??) is linearly added to the mixed spatial sources to generate a simulated fMRI data matrix of size 200 × 100 for each subject.\\
%%
We conducted simulation studies to evaluate the performance of the estimation and inference methods for the proposed strength of SC measure. We draw 300 simulated data sets for $n=20, 50$ subjects, under two different noise levels. First, we simulate the fMRI data based on the true source signal maps and their time courses. We define $q=2$ source IC maps, which are common for all subjects and consist of one $10 \times 10$ axial slice, for a total of $V=100$ voxels (see Figure 1). IC 1 represents a symmetric front-back network, while IC 2 represents a symmetric left-right network. The value at each voxel in these maps is based on the background noise ($x_1 \sim N(0,0.5$) for all voxels, plus the within-source intensity ($x_2=3$) and noise ($x_3 \sim N(0,0.1$)) for voxels within the component. After the IC maps are created, their temporal responses are adapted from real fMRI data with $T=200$ time points. We generate a $T \times V$ fMRI data matrix $Y_i$, for each subject $i$, according to the ICA equation: $Y_i=AS+e$, where $A$ is the $T \times q$ "mixing matrix" whose columns represent the time series for each IC, and $S$ is the $q \times V$ source matrix whose rows represent the IC maps. Gaussian background noise %with a variance of 1 %
is linearly added to the mixed spatial sources to generate a simulated fMRI data matrix of size $200 \times 100$ for each subject.\\
\\
\indent Once the subject-level fMRI data has been simulated, we can estimate the group-level IC maps using the GIFT method \citep{Calhoun2001}. In order to evaluate the strength of SC underlying each estimated IC, we must first simulate the probabilistic tractography SC results based on DTI data.  For each subject, we generate the $V \choose 2$ $\times 1$ matrix $\boldsymbol{N^*}$, whose elements ($N_{jk}$) represent the number of streams out of $N=20$ trials that connect each voxel pair ($j,k$), to simulate the results of a probabilistic tractography procedure. We use the model $\boldsymbol{N^*} \sim \text{MVN}(\boldsymbol{\mu}, \boldsymbol{\Sigma})$ to simulate this data, with  $\boldsymbol{\mu}$ and $\boldsymbol{\Sigma}$ defined as follows.  For the mean vector $\boldsymbol{\mu}=N\boldsymbol{p}$, where $\boldsymbol{p}$ is the vector of voxel pair connection probabilities. Voxel pairs outside of a component have a connection probability of 0.25, while voxel pairs inside IC 1 or 2 have connection probabilities of 0.5 and 0.75, respectively. The variance-covariance matrix $\boldsymbol{\Sigma}$ is defined based on the exponential semivariogram function, with parameters $c_0$ (nugget), $c_e$ (partial sill), and $a_e$ (range). We generate the SC data in $\boldsymbol{N^*}$ under both "low" ($c_0$=1, $c_e$=4, $a_e$=1) and "high" ($c_0$=2, $c_e$=5, $a_e$=1) noise conditions. In this way, we can simulate the SC data in $\boldsymbol{N^*}$, and evaluate the strength of SC underlying each IC. \\
\\
\indent Table 1 summarizes the results based on 300 simulation runs under the four different sampling configurations. In each setting, we estimate the strength of SC measure, $\hat{\theta_1}$ and $\hat{\theta_2}$, along with its variance and 95\% confidence intervals, based on both the theoretical variance term and the bootstrap standard error (using B=1000 bootstrap resamples). \\
\\
\indent We evaluate the bias of our sSC estimator by comparing the mean $\hat{\theta}$ to the true sSC value, $\theta$; there is very low bias in all simulation settings. We also assess the performance of our two candidate variance terms by comparing the estimated theoretical and bootstrap standard errors (SE) to the empirical SE, $\text{SD}(\hat{\theta})$. We note that the theoretical SE (based on semivariogram model fitting for $\boldsymbol{\hat{\Sigma}}$) tends to underestimate the variability of $\theta$, while the bootstrap SE performs fairly well. Finally, we compare coverage probabilities based on two types of 95\% confidence intervals (CIs): the Wald-type CI based on the theoretical SE, and the CI based on the bootstrap percentiles. The coverage probabilities from both types of CIs are fairly close to  $95\%$, although the bootstap CI tends to outperform the theoretical variance-based CI. \\
\\
\indent Because it requires estimating the large matrix $\boldsymbol{\hat{\Sigma}}$, calculation of the theoretical variance term $Var(\hat{\theta})$ poses a computational challenge, especially when the number of voxels $V$ is large. Even in this small-scale simulation study where $V=100$, $\boldsymbol{\hat{\Sigma}}$ has dimensions $100 \choose 2$ $\times$ $100 \choose 2$, or 4950 $\times$ 4950. The bootstrap variance estimator, on the other hand, is more computationally feasible since it avoids estimation of $\boldsymbol{\hat{\Sigma}}$, and shows good performance in our simulation studies. Thus, we recommend using the bootstrap method to conduct inference in real data applications where the number of voxels is large. \\
\\
\section{Application to real dataset}
\subsection{ssC analysis}
We apply our sSC method to an fMRI and DTI dataset of 20 subjects with major depressive disorder (MDD) and 20 healthy controls. Initially, we run a group ICA using only the 20 control subjects' fMRI data, since studies of MDD have shown resting state FC differences \citep{Northoff2011,Greicius2007,Veer2010}. We extract $q=15$ group-level IC maps, 9 of which appear to represent well-known resting state networks \citep{Smith2009,Laird2009} (See Figure 2). We create a thresholded white matter mask for each IC, consisting of about 900 voxels, for further exploration of the underlying structural connectivity. \\
\\
\indent To evaluate the SC distribution of each IC, we run a probabilistic tractography procedure using the voxels in the thresholded IC mask as seed locations. We initiate N=5000 streams from each seed voxel in the IC, and trace the streams as they pass through the brain. The results of this procedure give us $N_{jk}$, $\bar{N}_j$, and $\bar{N}_k$ for each voxel pair $j,k$ in IC $\ell$, which can be used to estimate the strength of SC measure, $\hat{\theta_\ell}$. We conduct inference for $\theta_\ell$ using the bootstrap SE term, rather than the theoretical $Var(\hat{\theta}_\ell)$ term for computational feasibility. \\
\\
\indent Table 2 shows the sSC results for the 9 estimated ICs from the control subject group. The $\hat{\theta}$ values are all fairly small, since our it was calculate on the voxel level, yet all ICs have strength of underlying SC significantly greater than 0. Since these IC maps all correspond to known resting state functional networks, it is not surprising that they demonstrate within-network strength of SC above baseline. IC 8 displays the highest mean $\hat{sSC}$, while IC 4 displays the lowest mean $\hat{sSC}$. To investigate this discrepancy, we plot the SC distribution for both of these ICs (see Figure 3), and see that IC 8 has a high degree of wihin-IC structural connectivity, while IC 4 has low SC within-IC relative to the rest of the brain. Permutation testing reveals the the strength of SC for these two ICs is significantly different (p$<$0.0001).\\
\\
\indent Next, we examine the difference in strength of SC for each IC between the control and MDD subject groups; the results are shown in Table 3. This table shows each group's mean $\hat{\theta}$ by IC, along with the bootstrap-based 95$\%$ confidence intervals and p-values (uncorrected) for the group difference in $\theta$; p-values are calculated based on both the bootstrap SE and permutation testing. The mean $\hat{\theta}$ values are very similar for the control and MDD groups, so it is not surprising that none of of the ICs show a significant between-group difference. This indicates that within the ICs that represent a normal resting state network, there is no substantial difference in the underlying structural connectivity between healthy controls and patients with MDD. \\ % This difference may be more evident in an analysis of a disease with known SC disruptions (e.g. Multiple Sclerosis, which is characterized by white matter atrophy).\\
\\
\indent Finally, we investigated whether strength SC informs the reliability of the functional networks estimated by data-driven methods like ICA. We plot the strength of SC measure $\hat{\theta_\ell}$ vs our proposed reliability index $R_\ell$ for each IC $\ell$, and find that these measures are positively associated (see Figure 4). That is, ICs with stronger underlying structural connectivity are more likely to be consistently estimated by the ICA algorithm. This suggests that we can leverage SC information from DTI data to inform the FC networks estimated from fMRI data. \\
\\
%%%%%%%%%%%%%%%%%%%%%%%%%%%%%%%%%%%%%%%%%%%%%%%%%%%%%%%
%%%%%%%%%%%%%%%%%%%%%%%%%%%%%%%%%%%%%%%%%%%%%%%%%%%%%%%
\section{Discussion}\label{conclusions}
We integrated information from the fMRI and DTI data modalities to calculate a novel statistical measure of the strength of SC (sSC) underlying a functional network. Our simulation studies and data application demonstrated the utility of the sSC measure, and found a positive association between sSC and the reliabilty of functional networks estimated by ICA.\\
%a paragraph about resting state FC???:
\\
There has been an enormous amount of research devoted to FC over the past several years, with substantial interest focusing on resting-state FC. A set of resting state networks (RSNs) have been consistently identified in these investigations (Smith et al., 2009; Damoiseaux et al., 2006; Laird et al., 2011), most prominently the default mode network (DMN; Buckner et al., 2008). While our work focused on resting-state connectivity, our proposed methods extend to other studies involving task-related fMRI.\\
\\
\indent Depression is a serious mental disorder affecting more than 20 million people in the US and roughly 121 million people worldwide, according to the WHO. Due to the complexity of this disorder and its varied forms, its mechanisms are not fully understood. Functional and structural neuroimaging play critical roles in advancing our knowledge about major depression and other mental disorders. Our proposed methods stand to make a significant impact by improving our understanding of the neural representations of MDD, concentrating largely on the functional and structural relationships between different brain regions. Our research may have a long-term impact that is even more profound, since our proposed methods may generalize to studies of brain connectivity for other mental and neurological disorders, as well as to treatment studies.\\
\\
\indent Preliminary resting-state studies of MDD have found functional connectivity changes in patients with MDD. Anand et al. (2011) found that untreated patients with MDD exhibited decreased connectivity  between the dorsal ACC and other areas, and that connectivity improved in the disrupted pathways after 6 weeks of treatment. Greicius et al. (2007) showed that patients with MDD had increased connectivity between certain regions of the default mode network (Zhang et al., 2010). Another recently-published study suggests a relationship between SC and FC in MDD (Kwaasteniet et al., 2013). Our method did not detect differences in the strength of SC underlying functional networks between patients with MDD and controls, but additional studies of resting-state fMRI and DTI in MDD are necessary to validate these findings.\\
%
%%\section*{Acknowledgements}
%%We'd like to thank Dr. Helen Mayberg's Lab in the Emory Department of Psychiatry for providing the data.

\singlespacing
\bibliographystyle{biom}
\bibliography{Topic1}

%\bibliographystyle{biom}
%\begin{thebibliography}{/Users/phebe/Documents/CBIS/Bibtex/Topic1.bib}
%\end{thebibliography}

\newpage
\section*{Appendix A: Figures and Tables}

\begin{center}
	\vskip5pt
	\includegraphics[width=4in]{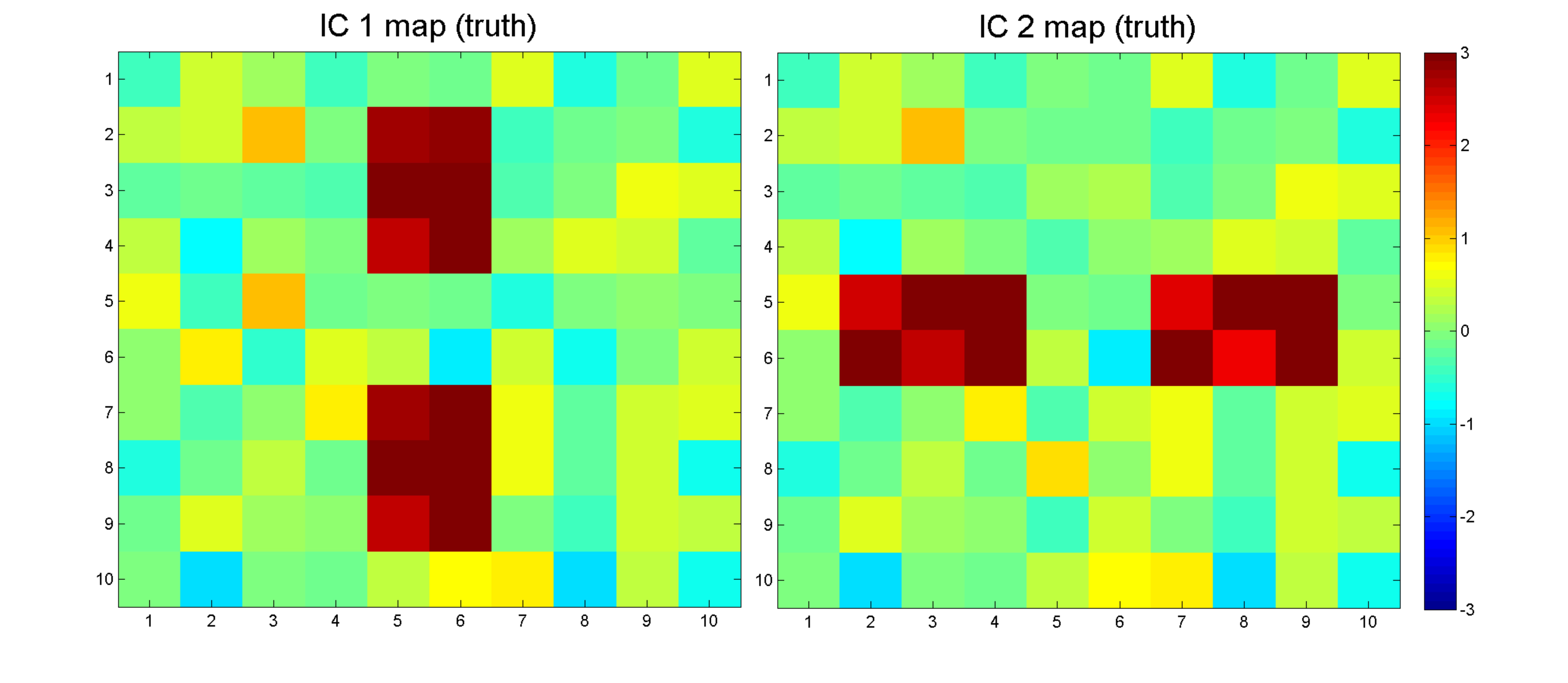}\\
\end{center}
\textbf{Figure 1}: True component maps specified for simulation testing.\\
\begin{center}
	\vskip5pt
	\includegraphics[width=3in]{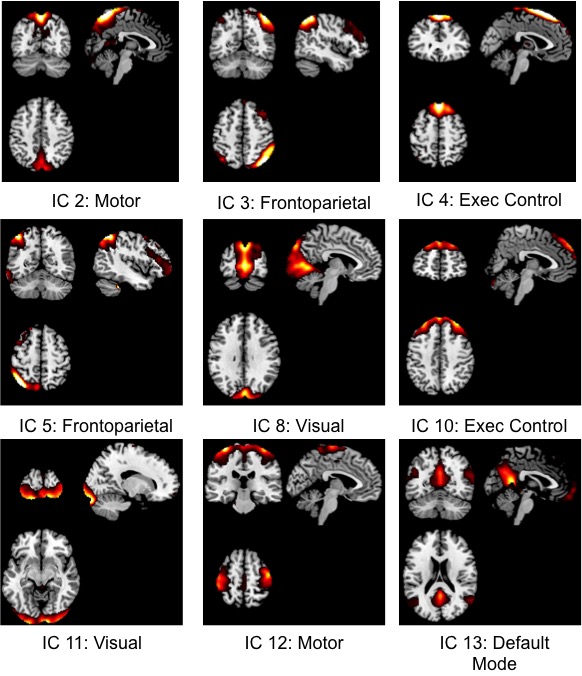}\\
\end{center}
\textbf{Figure 2}: Component maps estimated from group ICA of controls subjects' fMRI data. These maps correspond to resting state networks. \\

\newpage
 \thispagestyle{empty}
  	{\begin{tiny}
 \begin{landscape}
 	\begin{table}
 		
 		\caption {\it Results based on 300 simulation runs}
 		\begin{tabular}{ccllccccc}
 			\hline \hline
 			\\
 			& Sample   & Noise       &   $\theta$       &  $\hat{\theta}$  & Theoretical  & Bootstrap   & Cov. Prob. I$^{\ast}$   & Cov. Prob. II$^{\ast}$  \\
 			& size     & level     &          &  mean (SD)     & SE (SD)  & SE (SD)         & (Theoretical)    &   (Bootstrap)                  \\
 			\hline
 			\\
 			
 			IC 1 & 20  & Low  &   0.3077                 &0.3081  (0.0091)   & 0.0083 (0.00035) & 0.0093 (0.0016) &92.6 &94.3           \\
 			
 			&    & High  &   0.3077                 &0.3074  (0.0104)  &
 			0.0093 (0.00043) & 0.0105 (0.0018) &91.6  &94           \\
 			
 			& 50   & Low  &   0.3077                 &0.3084  (0.0061) &
 			0.0053 (0.00019)  & 0.0060 (0.00063)  &90.7   &93.7          \\
 			
 			&    &    High     &   0.3077              &0.3078  (0.0069)  &
 			0.0059 (0.00023) & 0.0068 (0.00071)  &91   &94.7                \\
 			
 			\hline
 			\\
 			
 			IC 2 & 20  & Low          &  0.64                 &0.6405  (0.0120)   & 0.0112 (0.00041)  & 0.0115 (0.0018)  &93.3  &93         \\
 			
 			&    & High           &  0.64                 &0.6389  (0.0134) &0.0126 (0.00040)  & 0.0127 (0.0020)  &94.6   &93.6         \\
 			
 			& 50   & Low          &  0.64                 &0.6409  (0.0080) &0.0071 (0.00019)  & 0.0074 (0.00073)  &90.7   &93         \\
 			
 			&    &    High       & 0.64                 &0.6394  (0.0088)  &
 			0.0080 (0.00017)  & 0.0082 (0.00081)   &93.7  &93             \\

 			\hline
 			\multicolumn{8}{l}{$^{\ast}$  {\footnotesize Coverage probabilities I: Based on Wald-type CI using theoretical SE.  \hspace{5mm} Coverage probabilities II: Bootstrap percentile confidence interval.}}\\
 		\end{tabular}
 		
 	\end{table}
 \end{landscape}
  \end{tiny}}
%\newpage
%
%\begin{center}
%	\vskip5pt
%	\includegraphics[width=5in]{ICs.jpg}\\
%\end{center}
% \textbf{Figure 2}: Component maps estimated from group ICA of controls subjects' fMRI data. These maps correspond to resting state networks. \\

\newpage

%add table 2 here!
 %\thispagestyle{empty}
 %\begin{landscape}
 	\begin{table}
 		
 		\caption {\it Results of hypothesis testing for controls}
 		\begin{tabular}{llllll}
 				%ccllccccc}
 			\hline \hline
 			\\
 			IC & & Mean($\hat{\theta}_\ell$) & SE$_{\text{boot}}$($\hat{\theta}_\ell$) & Bootstrap CI   & Bootstrap  \\
	 		   & &           &         &                & p-value  \\
 			\hline
 			\\
 			
 			%IC & & Mean(sSC) & SD(sSC) & Bootstrap CI  &   & Bootstrap p-value
 			2  & motor  & 0.0071 & 0.0013 & (0.0066, 0.0077) & $<$0.0001 \\
 			3  & FP     & 0.0081 & 0.0010 & (0.0077, 0.0086) & $<$0.0001 \\
 			\textbf{4}  & \textbf{EC}     & \textbf{0.0048} & 0.0004 & (0.0046, 0.0049) & $<$0.0001 \\
 			5  & FP     & 0.0077 & 0.0008 & (0.0074, 0.0081) & $<$0.0001 \\
 			\textbf{8}  & \textbf{visual} & \textbf{0.0098} & 0.0014 & (0.0092, 0.0103) & $<$0.0001 \\
 			10 & EC     & 0.0052 & 0.0008 & (0.0048, 0.0055) & $<$0.0001 \\ 			
 			11 & visual & 0.0085 & 0.0009 & (0.0082, 0.0089) & $<$0.0001 \\
 			12 & motor  & 0.0058 & 0.0005 & (0.0056, 0.0060) & $<$0.0001 \\
 			13 & DMN    & 0.0078 & 0.0016 & (0.0071, 0.0085) & $<$0.0001 \\  				
 			\hline
 			%\multicolumn{8}{l}{$^{\ast}$  {\footnotesize Coverage probabilities I: Based on Wald-type CI using theoretical SE.  \hspace{5mm} Coverage probabilities II: Bootstrap percentile confidence interval.}}\\
 		\end{tabular}
 		
 	\end{table}
 %\end{landscape}

   \begin{center}
   	\vskip5pt
   	\includegraphics[width=3in]{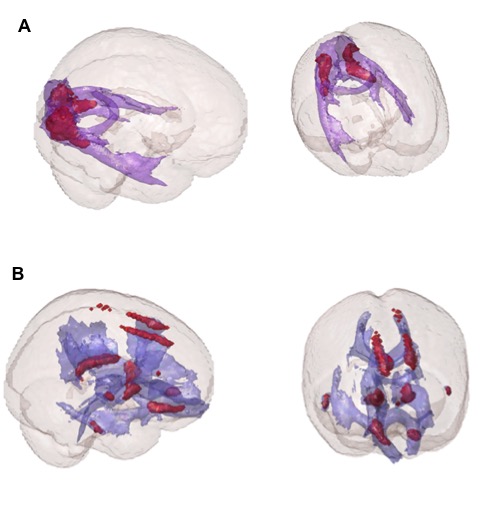}\\
   \end{center}
   \textbf{Figure 3}: Here, the IC seed area is shown in red, and its SC distribution to the rest of the brain is shown in blue. (A) IC 8 (a visual network) has the highest $\hat{\theta}$, which is confirmed by the high degree of SC within the IC. (B) IC 4 (an executive control network) has the lowest $\hat{\theta}$, which is supported by the low degree of SC within the IC relative to the rest of the brain. \\

% \newpage
% \begin{center}
% 	\vskip5pt
% 	\includegraphics[width=7in]{Table2.png}\\
% \end{center}

%\clearpage
%\newpage
%
%  \begin{center}
%  	\vskip5pt
%  	\includegraphics[width=5in]{SC_vis.jpg}\\
%  \end{center}
%  \textbf{Figure 3}: Here, the network seed area is shown in red, and its SC distribution to the rest of the brain is shown in blue. (A) The visual network (IC 8) has the highest $\hat{sSC}$, which is confirmed by the high degree of SC within the network. (B) The executive control network (IC 4)has the lowest $\hat{sSC}$, which is supported by the low degree of SC within the network \\
%
%
\newpage

%add table 3 here!
\thispagestyle{empty}
\begin{landscape}
\begin{table}
	
	\caption {\it Results of hypothesis testing for MDD vs. controls}
	\begin{tabular}{llllllllll}
		\hline \hline
		\\
		   & & \multicolumn{2}{c}{Controls}    & \multicolumn{2}{c}{MDD}&              &           &   \\
		IC & & Mean($\hat{\theta}$) & SE$_{\text{boot}}$($\hat{\theta}$) & Mean($\hat{\theta}$) & SE$_{\text{boot}}$($\hat{\theta}$) & Bootstrap CI & Bootstrap & Perm. test\\
		   & &           &         &           &         & (mdd - con)    & p-value   & p-value\\
		\hline
		\\
		%  & &           & Controls & & MDD     &        &                   &       &\\ 		
		2  & motor  & 0.0071 & 0.0013 & 0.0070  & 0.0012 & (-0.0009, 0.0005) & 0.702 & 0.698 \\
		3  & FP     & 0.0081 & 0.0010 & 0.0079  & 0.0012 & (-0.0009, 0.0005) & 0.525 & 0.562\\
		4  & EC     & 0.0048 & 0.0004 & 0.0047  & 0.0004 & (-0.0003, 0.0002) & 0.820 & 0.825\\
		5  & FP     & 0.0077 & 0.0008 & 0.0072  & 0.0010 & (-0.0011, -0.0001)& 0.018 & 0.054\\
		8  & visual & 0.0098 & 0.0014 & 0.0095  & 0.0016 & (-0.0012, 0.0007) & 0.639 & 0.608\\
		10 & EC     & 0.0052 & 0.0008 & 0.0053  & 0.0008 & (-0.0003, 0.0006) & 0.451 & 0.475\\ 			
		11 & visual & 0.0085 & 0.0009 & 0.0080  & 0.0008 & (-0.0010, 0.0000) & 0.084 & 0.077\\
		12 & motor  & 0.0058 & 0.0005 & 0.0060  & 0.0007 & (-0.0002, 0.0006) & 0.277 & 0.299\\
		13 & DMN    & 0.0078 & 0.0016 & 0.0073  & 0.0012 & (-0.0015, 0.0003) & 0.251 & 0.252 \\  				
		\hline
		%\multicolumn{8}{l}{$^{\ast}$  {\footnotesize Coverage probabilities I: Based on Wald-type CI using theoretical SE.  \hspace{5mm} Coverage probabilities II: Bootstrap percentile confidence interval.}}\\
	\end{tabular}
	
\end{table}
\end{landscape}

% \newpage
% \begin{center}
% 	\vskip5pt
% 	\includegraphics[width=7in]{Table3.png}\\
% \end{center}

\clearpage
\newpage

 \begin{center}
 	\vskip5pt
 	\includegraphics[width=5in]{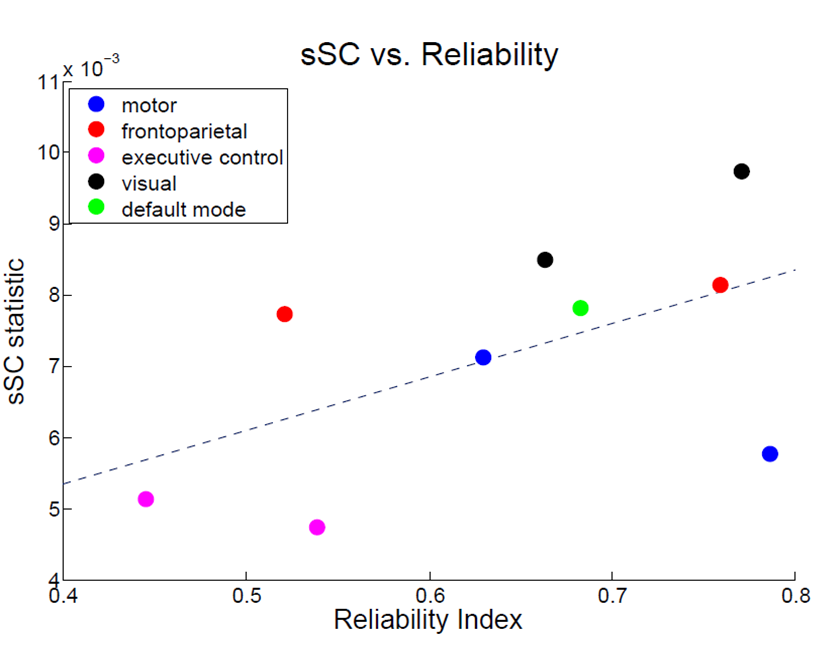}\\
 \end{center}
 \textbf{Figure 4}: Strength of SC is associated with IC reliability \\

\newpage
\section*{Appendix B}

\section*{$\hat{\theta}_{\ell}$ as a function of N*}

%create command that puts circles around numbers (better aligned than \textcircled{})
\newcommand*\circled[1]{\tikz[baseline=(char.base)]{
		\node[shape=circle,draw,inner sep=2pt] (char) {#1};}}

%\noindent The numerator and denominator of $\hat{\theta}_{\ell}$ can be written as linear combinations of \\ $\boldsymbol{ N^*}_{\substack{{V \choose 2}\times 1}}=\begin{bmatrix}
%N_{12}\\
%N_{13}\\
%\vdots \\
%N_{V-1,V}
%\end{bmatrix}$ \hskip10pt
%where $\boldsymbol{N^*} \sim \text{MVN}(\boldsymbol{\mu},\boldsymbol{\Sigma})$ \\
%\vskip10pt
%\noindent i.e. \hskip10pt	$\hat{\theta}_\ell=\dfrac{(\boldsymbol{C_{\ell}}-\boldsymbol{A})\boldsymbol{N^*}}{ b-\boldsymbol{AN^*}}$ \\

\noindent The numerator and denominator of $\hat{\theta}_{\ell}$ can be written as linear combinations $\boldsymbol{ N^*}$\\
	\begin{center}	$\hat{\theta}_\ell=\dfrac{(\boldsymbol{C_{\ell}}-\boldsymbol{A})\boldsymbol{N^*}}{ b-\boldsymbol{AN^*}}$
	\vskip10pt
	\noindent where \hskip5pt $\boldsymbol{A}=\dfrac{(V_{\ell}-1)}{2(V-1)} \displaystyle\sum_{\substack{j\in\Omega_{\ell}}}\boldsymbol{C_j}$ \hskip10pt and \hskip10pt $b=\dfrac{V_{\ell}(V_{\ell}-1)}{2}N$ \\
	\end{center}
\noindent Notation:  $\boldsymbol{C_{\ell}}$ and $\boldsymbol{C_j}$ are vectors of binary indicators, indicating which voxel pairs are members of $\Omega_{\ell}$ or involve voxel j, respectively. $V$ is the total number of voxels in the whole brain, and $V_{\ell}$ is the number of voxels in component $\ell$. $N$ is the maximum possible number of connections between a pair of voxels (i.e. the number of streams initiated from each voxel in a probabilistic tractography procedure).\\

\noindent \textbf{Proof:}\\
\\
$\hat{\theta}_{\ell} =\frac {\left. \displaystyle\sum_{\substack{j,k\in\Omega_{\ell}}}[N_{jk} -(\bar{N_j}+\bar{N_k})/2] \, \right\} \, \circled{1} }
{\left. \displaystyle\sum_{j,k\in\Omega_{\ell}}[N -(\bar{N_j}+\bar{N_k})/2] \, \right\} \, \circled{2}}$
\\
\vskip10pt
\noindent \textbf{$\hat{\theta}_{\ell}$ numerator:}

\begin{align*}
\circled{1} &=\displaystyle\sum_{\substack{j,k\in\Omega_{\ell}}}[N_{jk} -(\bar{N_j}+\bar{N_k})/2] \\
&= \underbrace{\displaystyle\sum_{\substack{j,k\in\Omega_{\ell}}}N_{jk}}_{\circled{3}} -\underbrace{\displaystyle\sum_{\substack{j,k\in\Omega_{\ell}}}[(\bar{N_j}+\bar{N_k})/2]}_{\circled{4}}
\end{align*}
\\
$\circled{3}=\displaystyle\sum_{\substack{j,k\in\Omega_{\ell}}}N_{jk} = \begin{bmatrix} 1&1&0&1&\hdots&0 \end{bmatrix}
\begin{bmatrix}N_{12} \\ N_{13} \\ \vdots \\ N_{V-1,V} \end{bmatrix} = \boldsymbol{C_{\ell} N^*} $
\\

Where $\boldsymbol{C_{\ell}}$ is a $1 \times {V \choose 2}$ vector of binary indicators, indicating which voxel pairs are members of $\Omega_{\ell}$.\\

\begin{align*}
\circled{4} &=\displaystyle\sum_{\substack{j,k\in\Omega_{\ell}}}[(\bar{N_j}+\bar{N_k})/2] \\
&= \frac{1}{2} \displaystyle\sum_{\substack{j,k\in\Omega_{\ell}}}\left[\frac{[N_{j1}+N_{j2}+\hdots +N_{jV}]}{V-1}+\frac{[N_{k1}+N_{k2}+\hdots +N_{kV}]}{V-1}\right] \\
&= \frac{1}{2}\displaystyle\sum_{\substack{j,k\in\Omega_{\ell}}}\left[\frac{\boldsymbol{C_j N^*}}{V-1}+\frac{\boldsymbol{C_k N^*}}{V-1}\right]
\\
&= \frac{1}{2(V-1)} \displaystyle\sum_{\substack{j,k\in\Omega_{\ell}}}[\boldsymbol{C_j}+\boldsymbol{C_k}]\boldsymbol{N^*}\\
&= \frac{V_{\ell}-1}{2(V-1)} \displaystyle\sum_{\substack{j \in\Omega_{\ell}}}\boldsymbol{C_j}\boldsymbol{N^*}
\end{align*}
where $\boldsymbol{C_j}$ is a $1 \times {V \choose 2}$ vector of binary indicators, indicating which voxel pairs include voxel $j$\\
\\
$\therefore \circled{1}=\left(\boldsymbol{C_{\ell}} -  \underbrace{\dfrac{V_{\ell}-1}{2(V-1)} \displaystyle\sum_{\substack{j \in\Omega_{\ell}}}\boldsymbol{C_j}}_{\boldsymbol{A}} \right)\boldsymbol{N^*} = (\boldsymbol{C_{\ell}}-\boldsymbol{A})\boldsymbol{N^*}$
%$\therefore \circled{1}=\left(\boldsymbol{C_{\ell}} -  \underbrace{ \dfrac{1}{2(V-1)} \displaystyle\sum_{\substack{j,k\in\Omega_{\ell}}}(\boldsymbol{C_j}+\boldsymbol{C_k})}_{\boldsymbol{A}} \right)\boldsymbol{N^*} = (\boldsymbol{C_{\ell}}-\boldsymbol{A})\boldsymbol{N^*}$
\\
\\
\vskip10pt
\textbf{$\hat{\theta}_{\ell}$ denominator:}
\begin{align*}
\circled{2} &=\displaystyle\sum_{\substack{j,k\in\Omega_{\ell}}}[N-(\bar{N_j}+\bar{N_k})/2] \\
&= \frac{V_{\ell}(V_{\ell}-1)}{2}N -\underbrace{\displaystyle\sum_{\substack{j,k\in\Omega_{\ell}}}[(\bar{N_j}+\bar{N_k})/2]}_{\circled{4}}
\end{align*}
\\
$\therefore \circled{2}= \underbrace{\frac{V_{\ell}(V_{\ell}-1)}{2}N}_{b} - \underbrace{\left(\frac{V_{\ell}-1}{2(V-1)} \displaystyle\sum_{\substack{j \in\Omega_{\ell}}} \boldsymbol{C_j} \right)}_{\boldsymbol{A}} \boldsymbol{N^*} = b-\boldsymbol{AN^*}$
%$\therefore \circled{2}= \underbrace{\dfrac{V_{\ell}(V_{\ell}-1)}{2}}_{b} - \underbrace{\left(\dfrac{1}{2(V-1)} \displaystyle\sum_{\substack{j,k\in\Omega_{\ell}}}(\boldsymbol{C_j}+\boldsymbol{C_k}) \right)}_{\boldsymbol{A}}\boldsymbol{N^*} = b-\boldsymbol{AN^*}$

\end{document}